\documentclass[11pt,onecolumn,conference]{IEEEtran}
\IEEEoverridecommandlockouts
\usepackage[letterpaper,margin=1.25in]{geometry}
\usepackage{microtype} 

\usepackage{cite}
\usepackage{amsmath,amssymb,amsfonts}
\usepackage{algorithmic}
\usepackage{graphicx}
\usepackage{textcomp}
\usepackage{xcolor}
\usepackage{bm}
\usepackage{pgfplots}
\usepackage{url}
\usepackage{verbatim}
\usepackage{cases}
\usepackage[ruled]{algorithm2e}

\def\BibTeX{{\rm B\kern-.05em{\sc i\kern-.025em b}\kern-.08em
    T\kern-.1667em\lower.7ex\hbox{E}\kern-.125emX}}

\begin{document}

\definecolor{mygray}{gray}{.9}
\newcommand{\blue}{\textcolor[rgb]{0.2265, 0.5859, 0.8633}}

\title{SCOPE: Sidelobe-Controlled Off-grid\\ Profile Estimation for Multiband\\ Multistatic Target Localization\\ in Upper Mid-Band ISAC Systems}

\author{\IEEEauthorblockN{Wenyu Huang, Nuria González-Prelcic\thanks{This material is based upon work partially supported by funds from the industry affiliate program at the Center for Wireless Communications (CWC) at UC San Diego.}\thanks{This paper has been accepted for presentation at the 2026 IEEE International Workshop on Signal Processing Advances in Wireless Communications (SPAWC). \copyright~2026 IEEE. Personal use of this material is permitted. Permission from IEEE must be obtained for all other uses, in any current or future media.}}
\IEEEauthorblockA{
Dept. Electrical and Computer Engineering, University of California, San Diego, USA\\
Email: \{wenyuhuang, ngprelcic\}@ucsd.edu}
}

\maketitle

\begin{abstract}
Multiband multistatic integrated sensing and communication (ISAC) in fragmented FR3 bands (7--24~GHz) enables high resolution localization via virtual wideband and spatial diversity. However, frequency anisotropy decorrelates target scattering across non-contiguous bands, while large inter-band frequency gaps generate severe grating lobes that create persistent ghost peaks. We propose sidelobe-controlled off-grid profile estimation (SCOPE), a robust localization algorithm that exploits multi-view consistency across distributed receivers and frequency bands to suppress grating-lobe ambiguities. At the transmitter, an \emph{iterative minimax precoder} suppresses out-of-region sidelobes to reduce false peaks in the coarse likelihood map. At the receiver, SCOPE employs profile likelihood with Top-K inhibition-based peak selection to avoid trapping in ghost basins, followed by derivative-free off-grid refinement. Simulations demonstrate that SCOPE achieves sub-meter localization with 90\% probability at $-5$\,dB SNR and 3\,mm root mean square error (RMSE) at 25\,dB SNR.
\end{abstract}


\section{Introduction}
\label{sec:intro}

Integrated sensing and communications (ISAC) is expected to be a core feature of 6G, enabling high-precision environmental awareness alongside data transmission. To achieve centimeter-level localization, operation is extending into the upper mid-band (FR3, 7--24 GHz)~\cite{gonzalez2025six}, which offers a balance between mmWave coverage limitations and Sub-6 GHz bandwidth shortage. However, FR3 spectrum is inherently fragmented, making the multiband operation essential. By aggregating non-contiguous component carriers (CCs), multiband systems can synthesize a large virtual bandwidth and significantly improve range resolution~\cite{gonzalez2024integrated}.

Despite this potential, multiband sensing at FR3 faces fundamental challenges absent in single band systems. As shown in~\cite{pegoraro2025toward}, target scattering properties, including radar cross section (RCS) and phase, vary significantly across widely separated bands, breaking the phase coherence required for traditional coherent combining. In addition, spectral fragmentation induces severe grating lobes, leading to range ambiguities.

Prior work has addressed these issues from different angles, but key limitations remain. Multistatic cooperative sensing has been explored to overcome single-node constraints~\cite{temiz2022improved, li2025multi}, where spatial diversity reduces false alarms and improves coverage. However, existing approaches largely focus on resource allocation or narrowband fusion, and do not address the aggregation of wideband, fragmented FR3 waveforms.

From a signal processing perspective, several works have explored coherent multiband fusion. \cite{wang2025compressed} proposed an alternating direction method of multipliers (ADMM) framework for compressed multiband sensing. \cite{zhang2024range} attempted to synthesize a virtual wideband spectrum by aligning DFT grids across carriers. However, their model relies on the idealized assumption that the channel is fully coherent across bands, ignoring the frequency anisotropy inherent to FR3 targets.
Recently, \cite{liu2025carrier} proposed restoring coherence by canceling RCS phase via conjugate multiplication across frequency bands. This approach, however, requires cross-band aggregation between widely separated bands (Sub-6 GHz and mmWave), inconsistent with typical intra-band carrier aggregation deployments, and fails under severe frequency anisotropy where the RCS amplitude and phase fluctuations become uncorrelated.
To address incoherence, \cite{pegoraro2025toward} proposed fusing non-contiguous band range profiles via product-based combining, which mitigates grating lobes in mono-static scenarios. However, the multiplicative fusion is vulnerable to band-specific deep fading, where a null in one band can suppress the target. It also fails to exploit spatial diversity in multistatic geometries.

In this paper, we address the limitations of strict coherent processing by proposing a robust framework that exploits the diversity inherent in the system rather than fighting the lack of coherence. We observe that while phase coherence is fragile in FR3, the geometric consistency of the true target is invariant and could be utilized.
Our main contributions are as follows:
\begin{itemize}
    \item We propose a robust transmit precoding scheme based on iterative minimax optimization to suppress out-of-band spectral leakage, purifying the interference landscape for the receiver.
    \item We develop the Sidelobe-Controlled Off-grid Profile Estimation (SCOPE) algorithm. By leveraging the \emph{double diversity}—frequency diversity from multiband operation to suppress grating lobes and spatial diversity from multistatic geometry to resolve angular ambiguities, SCOPE achieves robust high precision localization.
\end{itemize}

 \section{System Model}
\label{sec:preliminaries}

\begin{figure}[htbp]
  \centering
  \includegraphics[width=0.68\linewidth]{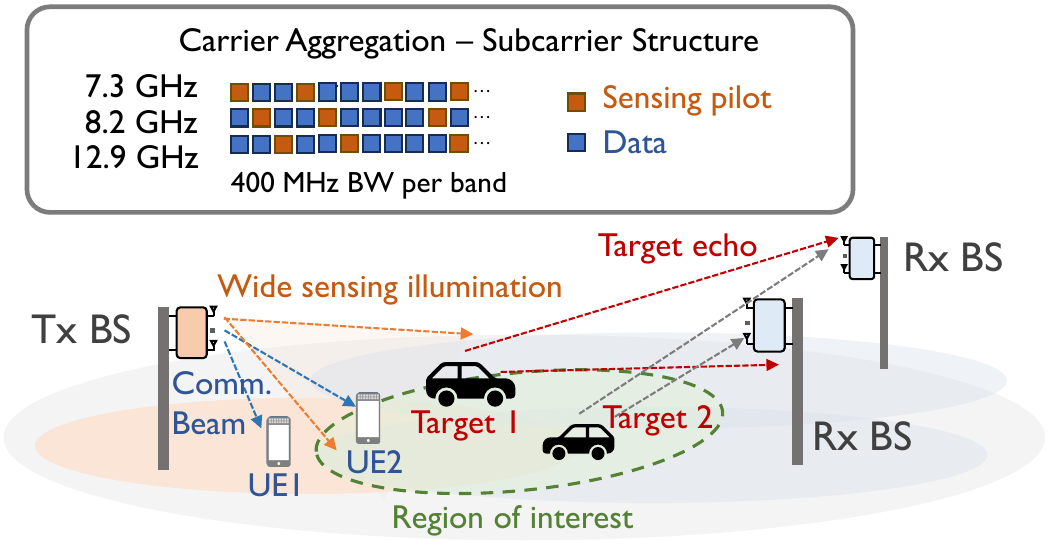}
  \caption{Overview of the multistatic multiband FR3 ISAC system architecture.}
  \label{fig:SystemArch}
\end{figure}

As shown in Fig.~\ref{fig:SystemArch}, we consider a multistatic ISAC network in the FR3 band, consisting of one transmit base station (Tx BS) and $R$ geographically distributed receive base stations (Rx BSs). The network serves downlink users while simultaneously localizing point-like targets in the environment. We denote the positions of $U$ such targets as $\{\mathbf{p}_u\}_{u=1}^U$, and focus on the single target case ($U=1$). The Tx BS is equipped with a uniform linear array (ULA) of $N_{\mathrm{tx}}$ antennas at a known position $\mathbf{p}_{\mathrm{tx}}\in\mathbb{R}^2$, and each Rx BS $r$ employs a ULA of $N_{\mathrm{rx}}$ antennas at a known position $\mathbf{p}_r\in\mathbb{R}^2$.

Let $\mathcal{M}_b$ denote the set of $M_b$ subcarriers in band $b$ allocated for sensing reference signals, where $f_{b,m} = f_c^{(b)} + m\Delta f$ is the frequency of subcarrier $m \in \mathcal{M}_b$. These $M_b$ subcarriers are randomly selected from the available resource grid, and the remaining subcarriers are allocated to communication.
On subcarrier $m\in\mathcal{M}_b$, the transmitted signal is
\begin{equation}
\label{eq:tx_signal}
    \mathbf{x}^{(b)}[m] = \mathbf{w}^{(b)}[m] s^{(b)}[m],
\end{equation}
where $s^{(b)}[m]\in\mathbb{C}$ is the known pilot symbol and $\mathbf{w}^{(b)}[m]\in\mathbb{C}^{N_{\mathrm{tx}}}$ is the precoder.
{Note that the precoder is digital and constant within a band, i.e. $\mathbf{w}^{(b)}[m]=\mathbf{w}^{(b)}$ for all $m\in\mathcal{M}_b$. Each band thus uses one $\mathbf{w}^{(b)}$, which is designed in Section~\ref{sec:method}.}

\label{subsec:channel_model}

Consider $U$ point-like targets located at $\{\mathbf{p}_u\}_{u=1}^U$. The frequency domain MIMO channel is modeled as the superposition of a static background and target reflections
\begin{equation}
\mathbf{H}^{(b)}_r[m] = \mathbf{H}^{(b)}_{\mathrm{bg},r}[m] + \mathbf{H}^{(b)}_{\mathrm{tgt},r}[m],
\end{equation}
where $\mathbf{H}^{(b)}_{\mathrm{bg},r}[m]$ captures the LoS path and static clutter. The target-induced component is given by
\begin{equation}
\label{eq:H_tgt}
\mathbf{H}^{(b)}_{\mathrm{tgt},r}[m] = \sum_{u=1}^{U} \beta^{(b)}_{r,u}\, \mathbf{a}_{\mathrm{rx}}(\theta_{r,u})\, \mathbf{a}_{\mathrm{tx}}^H(\phi_u)\, e^{-j2\pi f_{b,m}\tau_{r,u}},
\end{equation}
where $\mathbf{a}_{\mathrm{tx}}(\phi)\in\mathbb{C}^{N_{\mathrm{tx}}}$ and $\mathbf{a}_{\mathrm{rx}}(\theta)\in\mathbb{C}^{N_{\mathrm{rx}}}$ are the transmit and receive array steering vectors, $\phi_u$ and $\theta_{r,u}$ are the angle of departure (AoD) and angle of arrival (AoA) of target $u$, $\tau_{r,u}$ is the bistatic propagation delay, and $\beta^{(b)}_{r,u} = \sqrt{L_r^{(b)}}\,\sigma_u^{(b)}\, e^{j\psi_r^{(b)}}$ is the complex channel gain. The three factors capture path loss and frequency-dependent target scattering. $\psi_r^{(b)}\sim\mathcal{N}(0,\sigma_\psi^2)$ is the band-dependent phase offset at Rx-$r$. The path loss follows $L_r^{(b)}(\mathbf{p}) \propto (\lambda^{(b)})^2/ (d_{\mathrm{tx}}^2\,d_{\mathrm{rx},r}^2)$, where $\lambda^{(b)}$ is the carrier wavelength. $\sigma_u^{(b)} = a^{(b)}e^{j\varphi^{(b)}}$ is the anisotropic RCS. Its amplitude $a^{(b)}$ follows a log-normal distribution~\cite{richards2010principles}. The phase vector $\boldsymbol{\varphi}=[\varphi^{(1)},\ldots,\varphi^{(N_B)}]^T$ is modeled as a zero-mean correlated Gaussian, $\boldsymbol{\varphi}\sim\mathcal{N}(\mathbf{0},\mathbf{R}_{\varphi})$. The entries of $\mathbf{R}_{\varphi}$ decay as the frequency gap between bands grows. Nearby bands exhibit similar scattering behavior and remain strongly correlated, whereas bands far apart in frequency (e.g. 7\,GHz and 13\,GHz) are nearly uncorrelated. This is consistent with both the FR3 scattering measurements of~\cite{pegoraro2025toward} and the frequency-dependent cluster statistics in 3GPP TR~38.901~\cite{3gpp38901}.
The model represents the equivalent channel after practical timing alignment and carrier frequency offset (CFO) compensation. Following~\cite{pegoraro2024jump}, timing offset is removed by correlating consecutive channel impulse response (CIR) magnitude profiles and aligning the CIR in delay. CFO is mitigated by referencing each target path to a static path at the same Rx BS. As in this operation only the packet-varying common phase drift is removed, a link-dependent constant phase ambiguity remains and is absorbed into $\beta_r^{(b)}$.

At each Rx BS, after OFDM demodulation and removal of the known
pilot symbol $s^{(b)}[m]$, the effective channel vector is obtained as
\begin{equation}
    \hat{\mathbf{h}}^{(b)}_r[m] = \mathbf{H}^{(b)}_r[m]\mathbf{w}^{(b)}[m]
    + \tilde{\mathbf{n}}^{(b)}_r[m],
\end{equation}
where $\tilde{\mathbf{n}}^{(b)}_r[m] \in \mathbb{C}^{N_{\mathrm{rx}}}$
is the effective noise after pilot removal.
The background component $\hat{\mathbf{h}}^{(b)}_{\mathrm{bg},r}[m]$ varies slowly relative to the sensing interval and can be estimated from an initial reference snapshot.
Differential subtraction from the subsequent aligned channel estimates suppresses the static background contributions and yields an estimate of the target-induced component~\cite{liu2023integrated}. Accordingly,
\begin{equation}
    \hat{\mathbf{h}}^{(b)}_{\mathrm{tgt},r}[m]
    \triangleq
    \hat{\mathbf{h}}^{(b)}_r[m]-\hat{\mathbf{h}}^{(b)}_{\mathrm{bg},r}[m].
\end{equation}
Then the target-induced estimates across all $M_b$ subcarriers in band $b$ at Rx $r$ are stacked into
\begin{equation}
\label{eq:h_br}
    \hat{\mathbf{z}}^{(b)}_r = \mathrm{vec}\!\left(
    \left[\hat{\mathbf{h}}^{(b)}_{\mathrm{tgt},r}[m]\right]_{m \in \mathcal{M}_b}
    \right) \in \mathbb{C}^{M_b N_{\mathrm{rx}}}.
\end{equation}

\section{Proposed Framework: Robust Multiband Sensing}
\label{sec:method}

In this section, we propose a joint transmitter--receiver framework tailored for FR3 multistatic sensing. The main challenges are twofold: (i) non-contiguous carrier aggregation creates severe grating-lobe ambiguities, and (ii) FR3 target scattering is frequency-anisotropic, which makes conventional coherent phase combining unreliable.
Our approach exploits \emph{double diversity}: while carrier aggregation provides frequency diversity to separate true targets from grating lobes~\cite{liu2025carrier, wang2025compressed}, the distributed multistatic geometry provides spatial diversity for geometric disambiguation~\cite{li2025multi, temiz2022improved}.
Accordingly, the transmitter is designed with an \emph{iterative minimax precoder} to provide stable wide-area illumination, and the receiver uses the proposed SCOPE algorithm to noncoherently fuse multiband observations for target detection and position estimation.

\subsection{Transmitter Design: Iterative Minimax Sidelobe Suppression}
\label{subsec:precoder}

In our multistatic architecture, the transmitter provides spatially flat illumination over $\Theta_{\mathrm{ROI}}$ rather than a steered narrow beam. Since the distributed Rx BSs resolve angular ambiguities through multi-view geometric diversity, directional discrimination at the transmitter is not necessary and would bring latency overhead. More critically, narrow beam sweeping illuminates only one direction per frame, breaking the synchronous multistatic fusion.
In practice, flat illumination faces two challenges. A floodlight precoder formed by summing steering vectors over the ROI introduces spectral leakage and fixed sidelobes near the ROI edges due to the finite array aperture. Moreover, non-contiguous FR3 carrier aggregation already induces range-domain grating-lobe artifacts~\cite{pegoraro2025toward, wan2024ofdm}. Additional out-of-ROI leakage may illuminate strong clutter that folds back as ghost peaks, degrading localization accuracy. Target-location-aware designs (e.g., MVDR) are inapplicable since the target position is unknown \textit{a priori}, and direct convex beampattern optimization requires an external solver at each update. We therefore adopt a closed-form baseline precoder for flat ROI coverage, followed by iterative reweighted refinement to suppress the dominant out-of-ROI sidelobes.

\subsubsection*{Baseline Precoder}
We first construct a reference widebeam by summing the transmit steering vectors over a discretized angular ROI following~\cite{tagliaferri2025integrating}:
\begin{equation}
\mathbf{w}_{\mathrm{ref}}
= \eta \sum_{\varphi \in \Theta_{\mathrm{ROI}}}
\mathbf{a}_{\mathrm{tx}}(\varphi; f_{\mathrm{ref}}),
\label{eq:wref}
\end{equation}
where $\mathbf{a}_{\mathrm{tx}}(\varphi; f)\in \mathbb{C}^{N_{\mathrm{tx}}\times 1}$ is the transmit steering vector at angle $\varphi$ and frequency $f$, $f_{\mathrm{ref}}$ is the carrier used to build the reference beam, and $\eta$ is chosen such that $\|\mathbf{w}_{\mathrm{ref}}\|_2^2 = P_{\mathrm{tx}}$. Here, $P_{\mathrm{tx}}$ denotes the transmit power constraint. This construction provides broad illumination over $\Theta_{\mathrm{ROI}}$, but it does not explicitly control the sidelobes outside the ROI.

\subsubsection*{Iterative Sidelobe Suppression}
A rectangular target response is difficult to realize with a finite array and causes leakage near the ROI edges. We instead adopt a smooth target response:
\begin{equation}
d(\varphi) =
\begin{cases}
\cos^{\gamma}\!\left(\dfrac{\pi}{2}\dfrac{|\varphi|}{\varphi_{\max}}\right),
& |\varphi| \le \varphi_{\max}, \\[1mm]
0, & \text{otherwise},
\end{cases}
\label{eq:dphi}
\end{equation}
where $\Theta_{\mathrm{ROI}}=[-\varphi_{\max},\varphi_{\max}]$ and $\gamma>0$ controls the roll-off rate toward the ROI edges.

For each band $b$, we refine the precoder by fitting its beam response to the desired angular profile $d(\varphi)$ over a dense scan grid $\Theta_{\mathrm{scan}} \subset [-90^\circ,90^\circ]$ and a small set of representative frequency samples $\mathcal{F}_b$ within that band. Here, $\mathbf{a}_{\mathrm{tx}}^H(\varphi;f)\mathbf{w}$ denotes the transmit beam response produced by precoder $\mathbf{w}$ at angle $\varphi$ and frequency $f$. The design balances three objectives: matching the desired illumination inside the ROI, suppressing out-of-ROI leakage over the full visible sector, and limiting deviation from the baseline widebeam.
Accordingly, at iteration $t$, we solve
\begin{align}
\min_{\mathbf{w}\in\mathbb{C}^{N_{\mathrm{tx}}\times 1}}
& \sum_{f\in\mathcal{F}_b}\sum_{\varphi\in\Theta_{\mathrm{scan}}}
\omega^{(t)}(\varphi)
\left| \mathbf{a}_{\mathrm{tx}}^H(\varphi;f)\mathbf{w}
- d(\varphi) \right|^2 \nonumber\\
&\quad + \mu \|\mathbf{w}\|_2^2
+ \lambda \|\mathbf{w}-\mathbf{w}_{\mathrm{ref}}\|_2^2,
\label{eq:precoder_cost}
\end{align}
where the first term matches the actual beam response to the desired profile over angle and frequency, $\omega^{(t)}(\varphi)$ is an angle-dependent weight, $\mu>0$ is a small ridge term for numerical stability, and $\lambda>0$ controls how far the refined solution is allowed to deviate from the baseline precoder $\mathbf{w}_{\mathrm{ref}}$. The weights are initialized as $\omega^{(0)}(\varphi)=1$ inside the ROI and to a larger value outside the ROI, so that out-of-ROI leakage is penalized more strongly from the beginning.
{Since the fractional bandwidth within each CC is small, the beam pattern varies negligibly across subcarriers.
Thus, $\mathbf{w}$ is equivalent to the band-$b$ precoder $\mathbf{w}^{(b)}$ in Eq.~\eqref{eq:tx_signal}.
Optimizing over a few representative frequencies $\mathcal{F}_b$ is therefore sufficient to produce a stable beam over the entire band.}
Then defining
\begin{align}
\mathbf{R}_{\Sigma}^{(t)}
&= \sum_{f\in\mathcal{F}_b}\sum_{\varphi\in\Theta_{\mathrm{scan}}}
\omega^{(t)}(\varphi)\,
\mathbf{a}_{\mathrm{tx}}(\varphi;f)
\mathbf{a}_{\mathrm{tx}}^{H}(\varphi;f), \\
\mathbf{r}_{d}^{(t)}
&= \sum_{f\in\mathcal{F}_b}\sum_{\varphi\in\Theta_{\mathrm{scan}}}
\omega^{(t)}(\varphi)\, d(\varphi)\,
\mathbf{a}_{\mathrm{tx}}(\varphi;f),
\end{align}
the weighted least-squares update is
\begin{equation}
\widetilde{\mathbf{w}}^{(t)}
=
\left(
\mathbf{R}_{\Sigma}^{(t)} + (\mu+\lambda)\mathbf{I}
\right)^{-1}
\left(
\mathbf{r}_{d}^{(t)} + \lambda \mathbf{w}_{\mathrm{ref}}
\right).
\label{eq:wtilde_update}
\end{equation}
The resulting vector is then normalized to satisfy the transmit power constraint:
\begin{equation}
\mathbf{w}^{(t)}
=
\sqrt{P_{\mathrm{tx}}}\,
\frac{\widetilde{\mathbf{w}}^{(t)}}
{\|\widetilde{\mathbf{w}}^{(t)}\|_2}.
\label{eq:w_norm}
\end{equation}

To further reduce the strongest sidelobes, after each iteration we evaluate the average out-of-ROI beampattern over $\mathcal{F}_b$, identify the angles with the largest leakage, and increase their weights in the next iteration. This reweighting step approximates minimax suppression of the dominant sidelobes by repeatedly focusing the update on the current worst peaks, thereby creating a cleaner environment for the subsequent SCOPE algorithm.
Since $\mathbf{w}^{(b)}$ depends only on the array geometry and ROI definition, the precoder can be computed offline and stored as a compact codebook (order of kilobytes).

\subsection{Receiver Design: SCOPE Localization Algorithm}
\label{subsec:scope}
The receiver processes the global measurement vector $\hat{\mathbf{z}}_{\mathrm{all}}$ through the \emph{Sidelobe-Controlled Off-grid Profile Estimation} (SCOPE) algorithm in Algorithm~\ref{alg:scope}, which fuses observations from all $N_B$ bands and $R$ receivers without requiring strict cross-band phase coherence and estimates the target position.

\subsubsection{Multiband Profile Likelihood Formulation}
With the transmit illumination established, the receiver must
fuse $\hat{\mathbf{z}}^{(b)}_r$ to estimate the target
position.
Since the per-band gains $\beta_r^{(b)}$ carry independent phase offsets, coherent cross-band combining is unreliable. We instead treat them as nuisance parameters and profile them out.
The position estimation is formulated as a maximum likelihood (ML) problem.
For a hypothesized position $\mathbf{p}$, we construct a dictionary vector $\boldsymbol{\xi}_r^{(b)}(\mathbf{p}) \in \mathbb{C}^{M_b N_{\mathrm{rx}}}$ encoding the bistatic geometry of each link:
\begin{equation}
\label{eq:xi_def}
\boldsymbol{\xi}_r^{(b)}(\mathbf{p}) = \underbrace{g^{(b)}(\phi(\mathbf{p}))}_{\mathrm{Tx\ Gain}} \cdot \underbrace{\boldsymbol{\alpha}_r^{(b)}(\mathbf{p})}_{\mathrm{Delay\ Phase}} \otimes \underbrace{\mathbf{a}_{\mathrm{rx}}(\theta_r(\mathbf{p}))}_{\mathrm{Rx\ Steering}},
\end{equation}
where $g^{(b)}(\phi)\!\triangleq\!\mathbf{w}^{(b)H}\mathbf{a}_{\mathrm{tx}}(\phi)$ is the scalar Tx beamforming gain, and $\boldsymbol{\alpha}_r^{(b)}(\mathbf{p}) = [e^{-j2\pi f_{b,1}\tau_r(\mathbf{p})},\allowbreak \ldots,\allowbreak e^{-j2\pi f_{b,M_b}\tau_r(\mathbf{p})}]^{T}$ collects the bistatic delay phases across the $M_b$ subcarriers.
Assuming $\hat{\mathbf{z}}_{r}^{(b)} = \beta_r^{(b)}\boldsymbol{\xi}_r^{(b)}(\mathbf{p}) + \mathbf{n}$ and profiling out $\beta_r^{(b)}$, we can obtain the \emph{Profile Likelihood} cost:
\begin{equation}
\label{eq:profile_likelihood}
J(\mathbf{p}) = \sum_{b=1}^{N_B}\sum_{r=1}^{R} \frac{\left|\bigl(\boldsymbol{\xi}_r^{(b)}(\mathbf{p})\bigr)^H \hat{\mathbf{z}}_{r}^{(b)}\right|^2}{\|\boldsymbol{\xi}_r^{(b)}(\mathbf{p})\|_2^2 + \epsilon}.
\end{equation}
Double diversity is reflected here. The sum over $b$ provides \emph{frequency diversity}: grating lobes shift with carrier frequency, while the true target remains consistent across bands. The sum over $r$ provides \emph{spatial diversity}: geometrically consistent hypotheses are reinforced across viewpoints, whereas ghost peaks are suppressed.
{This has a deeper interpretation.
With $\beta_r^{(b)}$ unknown, the signal lies in the 1D subspace spanned by $\boldsymbol{\xi}_r^{(b)}(\mathbf{p})$.
Substituting the ML estimate of $\beta_r^{(b)}$ gives the projection statistic $|\boldsymbol{\xi}^H\hat{\mathbf{z}}|^2/\|\boldsymbol{\xi}\|^2$. This is actually the GLRT for a 1D subspace model and measures how strongly the observation aligns with the hypothesized target response.}

\subsubsection{Topological Inhibition Search Strategy}
Direct optimization of Eq.~\eqref{eq:profile_likelihood} is challenging due to the presence of numerous local maxima (ghosts). To address this, SCOPE adopts a hierarchical search strategy to ensure robust convergence, summarized in Algorithm~\ref{alg:scope} and detailed below.

\begin{algorithm}[t]
\caption{\textsc{SCOPE}: Sidelobe-Controlled Off-grid Profile Estimation}
\label{alg:scope}
\begin{algorithmic}[1]
\REQUIRE Observations $\{\hat{\mathbf{z}}^{(b)}_r\}$, precoders $\{\mathbf{w}^{(b)}\}$, ROI $\Omega$, source number $K$, suppression radius $r_{\mathrm{inh}}$
\ENSURE Estimated profile $\hat{\mathbf{p}}_{\mathrm{opt}}$
\STATE \textbf{Coarse mapping:} evaluate $\mathbf{M} \leftarrow \{ J(\mathbf{p}) : \mathbf{p} \in \mathcal{G}_c \}$
\STATE \textbf{Greedy selection:} $\mathcal{C} \leftarrow \emptyset$, $\mathcal{U} \leftarrow \mathcal{G}_c$
\FOR{$k = 1$ to $K$}
    \STATE $\mathbf{p}_k^{*} \leftarrow \arg\max_{\mathbf{p} \in \mathcal{U}} \mathbf{M}(\mathbf{p})$,\quad $\mathcal{C} \leftarrow \mathcal{C} \cup \{\mathbf{p}_k^{*}\}$
    \STATE $\mathcal{U} \leftarrow \{ \mathbf{p} \in \mathcal{U} : \|\mathbf{p} - \mathbf{p}_k^{*}\| \ge r_{\mathrm{inh}} \}$
\ENDFOR
\FOR{each $\mathbf{p}_k^{*} \in \mathcal{C}$}
    \STATE $\mathbf{p}_{k}^{(0)} \leftarrow \arg\max_{\mathbf{p} \in \mathcal{G}_{\mathrm{loc},k}} J(\mathbf{p})$
    \STATE $\hat{\mathbf{p}}_k \leftarrow \textsc{Nelder--Mead}(J(\cdot), \mathbf{p}_{k}^{(0)})$
\ENDFOR
\STATE \textbf{return} $\hat{\mathbf{p}}_{\mathrm{opt}} \leftarrow \arg\max_{k} J(\hat{\mathbf{p}}_k)$
\end{algorithmic}
\end{algorithm}

\noindent\textbf{Step 1) Coarse Global Mapping:}
We first evaluate $J(\mathbf{p})$ on a coarse grid $\mathcal{G}_c = \{\mathbf{p}_1, \dots, \mathbf{p}_G\}$ covering the ROI $\Omega$. This captures the global topology of the objective function.

\noindent\textbf{Step 2) Basin Diversity Selection:}
To avoid trapping in a single false basin (grating lobe), we extract $K$ candidates from \emph{distinct} topological basins. We define a set of unselected points $\mathcal{U}$, initially $\mathcal{G}_c$. In each step $k$, we select the peak $\mathbf{p}_k^*$ and then \emph{inhibit} its neighborhood:
\begin{align}
\mathbf{p}_k^* &= \arg\max_{\mathbf{p} \in \mathcal{U}} J(\mathbf{p}), \\
\mathcal{U} &\leftarrow \mathcal{U} \setminus \{ \mathbf{p} \in \mathcal{U} : \|\mathbf{p} - \mathbf{p}_k^*\| < r_{\mathrm{inh}} \}.
\end{align}
This guarantees that $\{\mathbf{p}_k^*\}_{k=1}^K$ represent diverse geometric hypotheses. {The inhibition enforces a minimum separation of $r_{\mathrm{inh}}$ between selected peaks, so they fall in different basins. This prevents picking several peaks from the same false basin and lets the search cover distinct grating lobe hypotheses.}

\noindent\textbf{Step 3) Local Grid Refinement:}
Around each coarse candidate $\mathbf{p}_k^*$, we generate a dense local grid $\mathcal{G}_{\mathrm{loc},k}$ to define a better initialization for the continuous solver:
\begin{equation}
\mathcal{G}_{\mathrm{loc},k} = \left\{ \mathbf{p}_k^* + \boldsymbol{\delta} \mid \boldsymbol{\delta} \in [-S, S]^2, \text{step} = \delta_{\text{fine}} \right\}.
\end{equation}
We then evaluate $J(\mathbf{p})$ on this sub-grid and select the best point $\mathbf{p}_{k}^{(0)}$ as the seed.

\noindent\textbf{Step 4) Derivative-Free Refinement:}
Computing gradients of $J(\mathbf{p})$ is costly because the precoder response has a complicated form.
We therefore use the derivative-free Nelder-Mead simplex method. It exploits the smooth local convexity of the main lobe to improve off-grid accuracy.
Out-of-$\Omega$ positions are rejected via a hard boundary (cost set to $-\infty$).

\noindent\textbf{Step 5) Global Consensus:}
The final estimate is the candidate with the highest refined likelihood: $\hat{\mathbf{p}} = \arg\max_k J(\hat{\mathbf{p}}_k)$.

\section{Numerical Results}\label{sec:results}

We evaluate the proposed framework in a multistatic FR3 ISAC network with one Tx BS and two distributed Rx BSs. The Tx uses a 32-element ULA at $\mathbf{p}_{\mathrm{tx}} = [0, 0]^T$ m, while two Rx BSs with 8-element ULAs are located at $\mathbf{p}_{\mathrm{rx},1} = [50, 0]^T$ m and $\mathbf{p}_{\mathrm{rx},2} = [25, 40]^T$ m, forming a triangular geometry over the position search region $\Omega = [15, 45] \times [-15, 15]$ m. We set $P_{\rm tx} = N_{\rm tx}$ and $\gamma = 1.5$.
{The transmit precoder illuminates the angular ROI $\Theta_{\mathrm{ROI}}=[-30^\circ,30^\circ]$. The target is placed at random in $\Omega$ each trial.
Localization is easier within a smaller ROI, so for fair comparison, we fix this ROI for all results.}

The system operates over $N_B=3$ non-contiguous CCs at $f_c \in \{7.3, 8.2, 12.9\}$ GHz, each with 400 MHz bandwidth and subcarrier spacing $\Delta f = 120$ kHz. $M_b = 64$ subcarriers per band are used for sensing. The anisotropic RCS is modeled as $\sigma^{(b)} = a^{(b)} e^{j\varphi^{(b)}}$ with $a^{(b)}=10^{x^{(b)}/20}$, $x^{(b)}\sim\mathcal{N}(0,3^2)$ dB, and phase correlation $\mathbf{R}_{\varphi}$ with $\rho_{12}=0.85$, $\rho_{13}=\rho_{23}=0.10$~\cite{pegoraro2025toward}. Residual synchronization errors are modeled as $\psi_r^{(b)}\sim\mathcal{N}(0,(10^\circ)^2)$.

We compare SCOPE (Top-$K=6$) with MF-COH, MF-PROD~\cite{pegoraro2025toward}, and single-node SCOPE (Tx–Rx1 only).
{The single-node baseline uses only the Tx--Rx1 link. Any improvement of full SCOPE over it therefore comes from the second receiver, isolating the gain from spatial diversity.}

\subsection{Performance Analysis}

\subsubsection{Localization Accuracy (RMSE)}

\begin{figure}[t]
    \centering
    \includegraphics[width=0.74\linewidth]{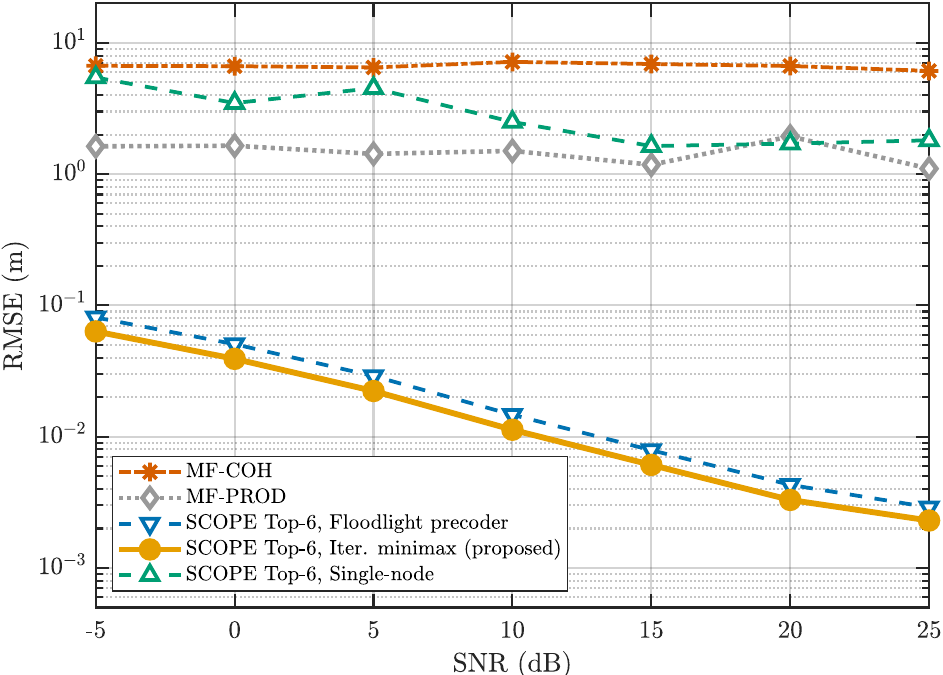}
    \caption{RMSE of localization error versus SNR.}
    \label{fig:rmse}
\end{figure}

Fig.~\ref{fig:rmse} shows RMSE versus SNR. MF-COH exhibits an error floor above 1\,m due to inconsistent cross-band phase caused by FR3 anisotropy.
MF-PROD avoids phase sensitivity but is vulnerable to deep fades. A single weak band suppresses the product and increases noise sensitivity at low SNR.
SCOPE mitigates both effects via noncoherent accumulation across bands and receivers.

The proposed precoder further improves performance, especially at low and medium SNR. It reduces illumination outside the ROI and provides better initialization.
SCOPE achieves millimeter-level accuracy ($\approx 3$\,mm at 25\,dB).

\subsubsection{Robustness and Outliers (CDF)}

\begin{figure}[t]
    \centering
    \includegraphics[width=0.72\linewidth]{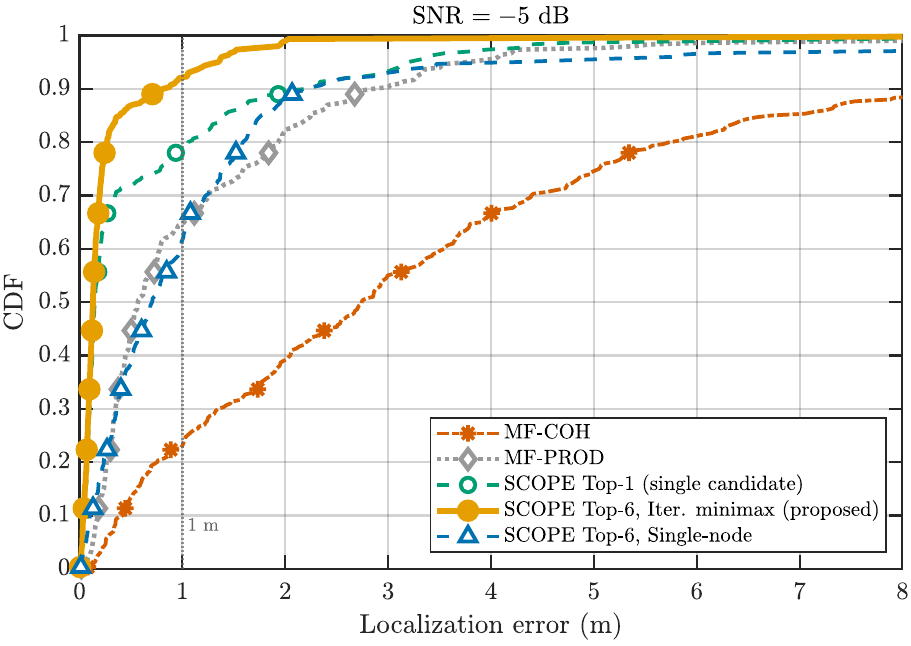}
    \caption{CDF of localization error at SNR = $-5$\,dB.}
    \label{fig:cdf}
\end{figure}

Fig.~\ref{fig:cdf} shows the CDF at SNR$=-5$\,dB. Single-node saturates near CDF$=0.8$, indicating $\sim$20\% failures due to bistatic ambiguity. SCOPE Top-1 reduces this but retains $\sim$10\% outliers.

SCOPE Top-$K$ resolves this by exploring multiple candidate basins, achieving $>$98\% sub-meter accuracy. {The precoder lowers the amplitude of ghost peaks, so the true peak stands out and initialization improves. This reduces average RMSE but cannot remove the ghost basins.
Outliers still occur when a ghost peak has higher likelihood than the true target.
Resolving that requires spatial diversity from multiple receivers.}

\subsection{Complexity Analysis}
The cost of evaluating $J(\mathbf{p})$ at a single grid point is $C_{\mathrm{eval}} \approx \mathcal{O}(N_B \cdot R \cdot M_b \cdot N_{\mathrm{rx}})$, dominated by inner products over subcarriers and receive links. The coarse grid search over $G$ candidate positions costs $\mathcal{O}(G \cdot C_{\mathrm{eval}})$ in total.
Since all $G$ evaluations are independent, this stage is fully parallelizable and maps directly to vectorized batch computation on a GPU.
The Top-$K$ refinement stage evaluates a dense local sub-grid around each of the $K$ selected candidates.
The grid size is $|\mathcal{G}_{\mathrm{loc},k}|$, as defined in Algorithm~1. Each candidate is then refined by derivative-free Nelder-Mead optimization.
This gives a refinement cost of
$\mathcal{O}(K \cdot (|\mathcal{G}_{\mathrm{loc},k}| + \upsilon \cdot N_{\mathrm{NM}}) \cdot C_{\mathrm{eval}})$,
where $\upsilon$ is the number of local seeds for each candidate and $N_{\mathrm{NM}}$ is the number of function evaluations for each Nelder-Mead run.
Since $K$, $\upsilon$, and $N_{\mathrm{NM}}$ are all small constants relative to $G$, the sequential refinement stage adds acceptable overhead. The dominant coarse stage can be accelerated on existing parallel hardware, while the lightweight refinement stage introduces only marginal additional cost.

\section{Conclusions}\label{sec:conclusion}
We proposed a high precision localization framework for fragmented FR3 ISAC systems. An iterative minimax precoder suppresses out-of-ROI sidelobes to purify the coarse likelihood map, while SCOPE exploits frequency and spatial diversity to resolve grating lobe ambiguities without cross band phase coherence. Simulations confirm sub-meter accuracy at $-5$\,dB SNR and millimeter level RMSE at high SNR. Future work will extend to multiple and moving targets scenarios.

	\bibliographystyle{IEEEtran}
	\bibliography{IEEEabrv,ref}

\end{document}